\documentclass[sigconf,natbib]{acmart}
\usepackage{multirow,multicol}
\usepackage{color,tabularx, colortbl,amsmath,bm,arydshln} %amssymb
\usepackage{enumitem,kantlipsum,arydshln}
\usepackage{booktabs,url,tcolorbox,mathtools} 
\usepackage{amsmath}
\usepackage{tabularx}
\usepackage{graphicx}
\usepackage{filecontents}
\usepackage{multirow,multicol}
\usepackage{enumitem,kantlipsum,arydshln}
\usepackage{booktabs,url,tcolorbox,mathtools} % For formal tables
\usepackage[stable]{footmisc}
\usepackage{newtxtext,newtxmath}
\usepackage{caption}
\usepackage{float}

\renewcommand\footnotetextcopyrightpermission[1]{} % removes footnote with conference information in first column
\newcommand\blfootnote[1]{%
  \begingroup
  \renewcommand\thefootnote{}\footnote{#1}%
  \addtocounter{footnote}{-1}%
  \endgroup
}

\usepackage{titlesec}

\begin{document}

\title{Retrieving Supporting Evidence for Generative Question Answering}

%\author{\em Anonymous author(s)}

\author{Siqing Huo}
\affiliation{%
  \institution{University of Waterloo}
  \city{Waterloo}
  \state{Ontario}
  \country{Canada}}
\email{siqing.huo.canada@gmail.com}

\author{Negar Arabzadeh}
\affiliation{%
  \institution{University of Waterloo}
  \city{Waterloo}
  \state{Ontario}
  \country{Canada}}
\email{narabzad@uwaterloo.ca}

\author{Charles L. A. Clarke}
\affiliation{%
  \institution{University of Waterloo}
  \city{Waterloo}
  \state{Ontario}
  \country{Canada}}
\email{claclark@gmail.com}
\begin{abstract}

Current large language models (LLMs) can exhibit near-human levels of performance on many natural language-based tasks, including open-domain question answering. Unfortunately, at this time, they also convincingly hallucinate incorrect answers, so that responses to questions must be verified against external sources before they can be accepted at face value. In this paper, we report two simple experiments to automatically validate generated answers against a corpus. We base our experiments on questions and passages from the MS MARCO (V1) test collection, and a retrieval pipeline consisting of sparse retrieval, dense retrieval and neural rerankers. In the first experiment, we validate the generated answer in its entirety. After presenting a question to an LLM and receiving a generated answer, we query the corpus with the combination of the question $+$ generated answer. We then present the LLM with the combination of the question $+$ generated answer $+$ retrieved answer, prompting it to indicate if the generated answer can be supported by the retrieved answer. In the second experiment, we consider the generated answer at a more granular level, prompting the LLM to extract a list of factual statements from the answer and verifying each statement separately. We query the corpus with each factual statement and then present the LLM with the statement and the corresponding retrieved evidence. The LLM is prompted to indicate if the statement can be supported and make necessary edits using the retrieved material. With an accuracy of over 80\%, we find that an LLM is capable of verifying its generated answer when a corpus of supporting material is provided. However, manual assessment of a random sample of questions reveals that incorrect generated answers are missed by this verification process. While this verification process can reduce hallucinations, it can not entirely eliminate them.
\blfootnote{
A preliminary version of this work appeared at the REML 2023 workshop. The final version will appear at SIGIR-AP 2023.}

\end{abstract}

\begin{CCSXML}
<ccs2012>
   <concept>
       <concept_id>10002951.10003317</concept_id>
       <concept_desc>Information systems~Information retrieval</concept_desc>
       <concept_significance>500</concept_significance>
       </concept>
   <concept>
       <concept_id>10010147.10010178.10010179.10010182</concept_id>
       <concept_desc>Computing methodologies~Natural language generation</concept_desc>
       <concept_significance>500</concept_significance>
       </concept>
 </ccs2012>
\end{CCSXML}

\ccsdesc[500]{Information systems~Information retrieval}
\ccsdesc[500]{Computing methodologies~Natural language generation}

%% the work being presented. Separate the keywords with commas. <TODO:>
\keywords{}

\maketitle
\pagestyle{plain} 

\section{Introduction}
There has been rapid progress in the field of Natural Language Processing due to recent advancements in transformer-based large language model (LLM)s~\cite{devlin2019bert,gpt3,lewis2019bart,radford2018improving,transformer}. These LLMs have produced substantial improvements in text generation tasks such as question answering, summarization, and machine translation~\cite{clinchant2019use,karpukhin2020dense,liu2019fine,miller2019leveraging,wang2019multi,yang2019end,zhu2020incorporating}. However, despite the excitement created by these improvements, the LLMs may confidently and convincingly generate hallucinated results~\cite{chatgpt_hallucination,NLG_hallucination}. Avoiding hallucinations is particularly important when LLM generated text is presented directly to users, especially in critical circumstances, such as health and medicine~\cite{chatbot-hallucination}.

Current LLMs lack the ability to self-detect hallucinations in generated texts as they do not have access to an external source of knowledge~\cite{chatgpt_hallucination}. On the other hand, information retrieval methods have been long studied and are now capable of rapidly locating the top documents relevant to queries from arbitrarily large text corpora~\cite{IRSurvey}. Attribution~\cite{attributedQA, attributedNLG, attributedLLM} focuses on connecting generated texts to supporting evidence to make them more trustworthy. Retrieval-augmented generation approaches~\cite{retrieval-aug-tagging, retrieval-aug-nlp} attempt to ensure the reliability of generated texts by conditioning the LLM's generation on retrieved material. However, such approaches still suffer from hallucination and cannot guarantee attribution. The LLM may make claims not found in the retrieved material~\cite{retrieval-aug-invent} or contradictory to the retrieved material~\cite{retrieval-aug-false}. 

As opposed to retrieval-augmented generation that performs retrieval before generation, more recent works such as RARR~\cite{RARR} proposed to perform retrieval after generation. The proposed framework suggests examining the produced text and making edits to align it with the gathered evidence, while maintaining the overall structure of the original text in case of any contradictions. Since our focus is on self-detecting and correcting hallucinations, we also perform retrieval after generation. However, unlike RARR which makes uses of few-shot prompting~\cite{brown2020language} and external query-document relevance model, our experiments use nothing else besides the LLM itself and a retrieval pipeline. Substantial prompt engineering is not needed. 

The LLM's generated texts is often more than just a single and atomic factual claim, it can be helpful to decompose a piece of generated text into a series of factual claims. Many previous works have studied decomposing long piece of text into atomic factual claims~\cite{fact-decompose-2023-01, FactScore, fact-decompose-2022}. While we took inspiration from these works, we use solely the LLM itself to achieve such decomposition with no training, no human intervention, and minimal prompt engineering.

In this paper, we investigate the ability for LLMs to self-detect hallucinations by confirming its generated responses against an external corpus. More specifically, we experimentally test the degree to which an LLM hallucinates answers when performing an open-domain, general question-answering task, and whether it can automatically verify its responses when presented with a dataset containing known correct answers, with the help of retrieval methods. Our experiments include manual checks of comparisons made by the LLM. These experiments demonstrate that the LLM can correctly detect its own hallucinations in a majority of cases (an accuracy of over 80\%), with the help of retrieval methods. However, while our verification process can reduce hallucinations, it can not entirely eliminate them. One should still be cautious when depending on LLM-generated answers, especially in critical circumstances.

\section{Experimental Setup}
In our experiments, we choose \verb|gpt-3.5-turbo| as the LLM representative with the temperature set to 0, consistent with OpenAI recommendations for classification tasks. We used the MS MARCO (V1) passage collection\footnote{https://github.com/microsoft/MSMARCO-Passage-Ranking} \cite{nguyen2016ms} for questions and answer validation. MS MARCO is a large-scale dataset with over 8 million passages for the development and evaluation of machine reading comprehension models. MS MARCO is accompanied by sparsely labeled queries as its training set, development set and test set. In this paper, we run experiments on the 6980 questions in the MS MARCO (V1) small development set. 

We run our set of experiments with two different retrieval methods. 
As the first retriever, we employ the Okapi BM25~\cite{bm25} ranking function, which  is a well-known and widely-used baseline retrieval method. For the BM25 function parameters, $k1$ is set to 0.82 and $b$ is set to 0.68, which are standard values tuned for the MS MARCO passage retrieval task by grid search. Since BM25 requires exact matching between query terms and document terms, we speculate that it may perform well for answer verification by providing support for the terms used in the generated answer.

The second retrieval method we adopt for our experiments is a more modern neural retrieval method that emphasizes the quality of the retrieved passages over retrieval efficiency. The pipeline comprises an initial retrieval stage followed by a reranking stage. For the retrieval stage, we employ a combined pool of sparse and dense retrieval. We use SPLADE~\cite{Splade} as the sparse retrieval method, and ANCE~\cite{ANCE} as the dense retrieval method. Both retrieval methods are shown to be highly effective \cite{thakur2021beir,yu2021few}. We pool the top 100 documents retrieved by both retrieval methods. For the reranking stage, we use a combination of MonoT5 and DuoT5 neural rerankers~\cite{MonoDuo}. We used MonoT5 to rerank the pooled documents from the retrieval stage, and we use DuoT5 to rerank the top 10 documents selected by MonoT5. We select this multi-stage neural retrieval stack (SPLADE$+$ANCE$+$MonoT5$+$DuoT5) as similar approaches have shown excellent performance on the MS MARCO passage ranking task. Executing our implementation of this neural retrieval pipeline on the MS MARCO (V1) small development set achieves a MRR@10 of 0.40. All the aforementioned methods are implemented using the Pyserini toolkit with default parameters\footnote{\url{https://github.com/castorini/pyserini/}}~\cite{pyserini}.

\section{Experiment 1}
Our first experiment is essentially the simplest method that we could envision for employing LLMs to self-verify against retrieved passages. We first prompt the LLM to answer the question. We then combine its generated answer with the original question and use the result to query a corpus of passages expected to contain supporting evidence. In order for the LLM to self-detect hallucinations, we then present the question, the generated answer, and the potential evidence to the LLM, prompting it to determine if the evidence supports the answer. 

\subsection{Methodology}

\begin{figure}[]
  \centering
  \scalebox{0.9}{
  \includegraphics[width=\linewidth]{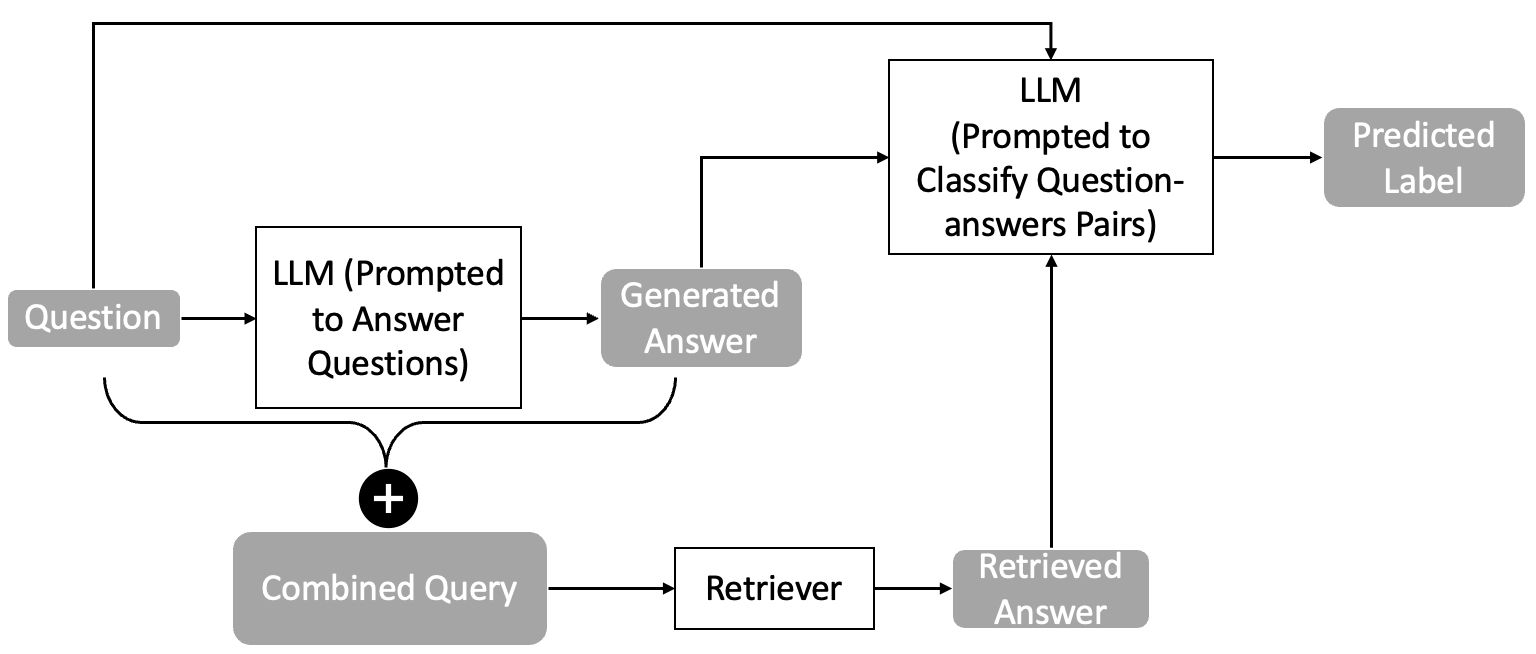}
  }
  \caption{Self-detecting hallucination in LLMs.}
  \label{fig:methodology_overview}
\end{figure}

\begin{figure}[]
  \centering
  \scalebox{0.8}{
  \includegraphics[width=\linewidth]{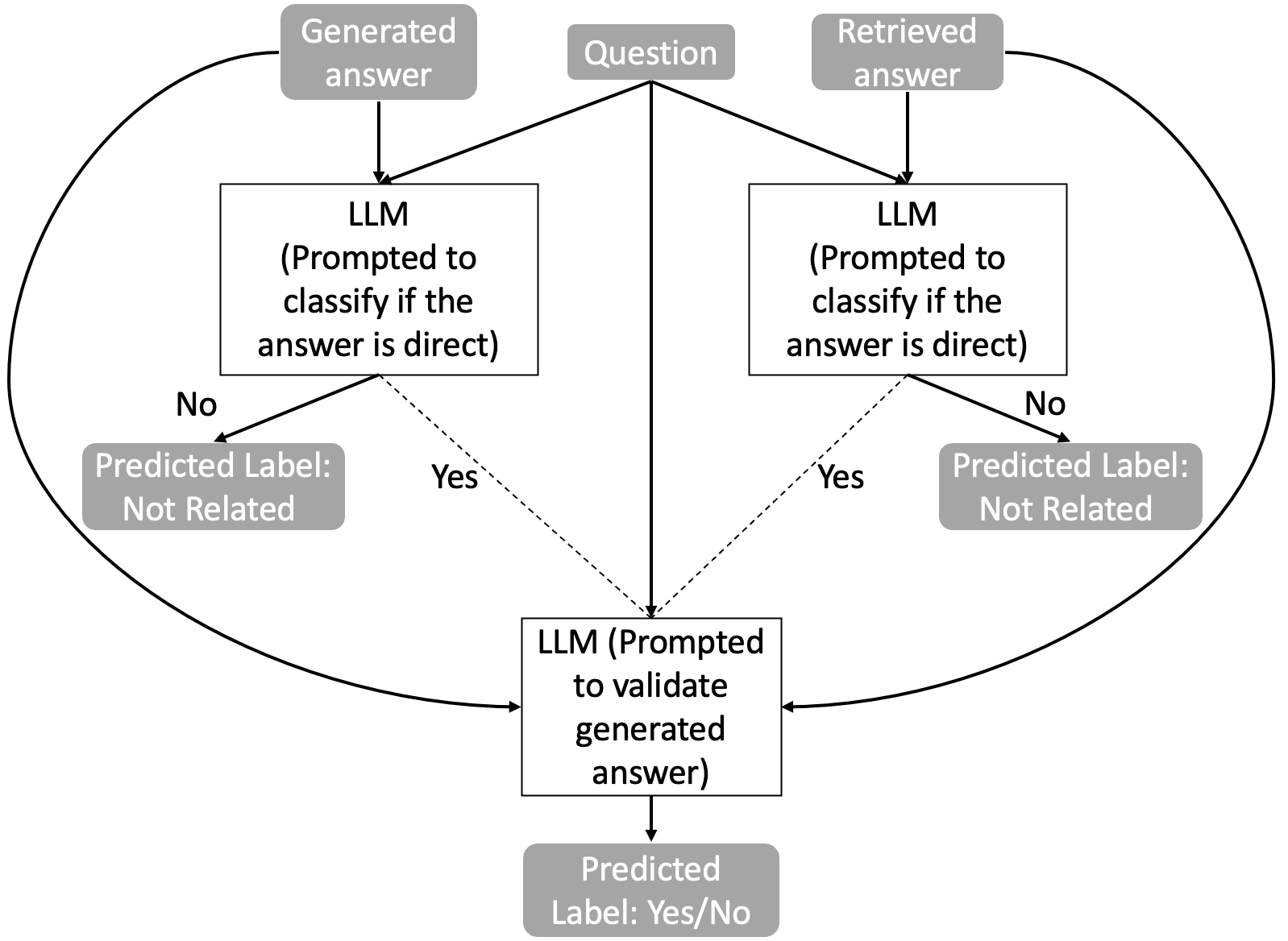}
  }
  \caption{Stepped classification of a question-answers pair.}
  \label{fig:stepped_classification}
\end{figure}

\begin{figure}[t]
\begin{tcolorbox}[colback=gray!5!white,colframe=gray!75!black]
\small{
  You are an expert in this field. Please answer the question as simply and concisely as possible.
  \newline
  \newline
  Question: \{query\}
  
  Answer:
  }
\end{tcolorbox}
\caption{Prompt for answering question.}
\label{fig:answer_prompt}
\end{figure}

\begin{figure}[t]
\begin{tcolorbox}[colback=gray!5!white,colframe=gray!75!black]
\small{
  I want you to act as an assessor of the answer. You will be given a question and an answer, and you need to determine whether the answer directly answers the question. Examples of non-direct answers would be claiming it does not know or does not have enough information, and provide some alternative ways to find answers. Also note that if an answer claims that the question itself is wrong, it also is a form of direct answer. Your response should be `Yes' if the answer actually answers the question, ad `No' if the answer does not actually answer the question. Please also include a short and concise explanation of your classification.
  \newline
  \newline
  Question: \{query\}
  
  Answer: \{answer\}
  }
\end{tcolorbox}
\caption{Prompt for assessing whether the answer directly addresses the question.}
\label{fig:check_direct_answer_prompt}
\end{figure}

\begin{figure}[t]
\begin{tcolorbox}[colback=gray!5!white,colframe=gray!75!black]
\small{
  I want you to act as an assessor of the answer. You will be provided with a question, an answer, and relevant evidence. Your task is to assess whether the evidence provided supports the given answer. If the evidence supports the answer, reply with a 'Yes'. Otherwise, reply with a 'No'. Please also include a short and concise explanation of your classification.
  \newline
  \newline
  Question: \{query\}
  
  Answer: \{LLM answer\}
  
  Evidence: \{Retrieved answer\}
  }
\end{tcolorbox}
\caption{Prompt for validating generated answer.}
\label{fig:validate_generated_answer_prompt}
\end{figure}

\begin{figure}[t]
\begin{tcolorbox}[colback=gray!5!white,colframe=gray!75!black]
\small{
  I want you to act as a question-based summarizer for a set of passages. Given a question and a passage containing answer to the question, your task is to provide a clear and concise summary of the passage that directly answer the question and contain minimal extra information. Your summary should be easy to understand and accurately represent the passage. Keep in mind that your summary should be objective and avoid including personal opinions or biases. If the passage does not answer, simply reply with `No Answer', otherwise reply with just the summary itself and nothing else.
  \newline
  \newline
  Question: \{query\}
  
  Passage 1: \{passage1\}

  Passage 2: \{passage2\}

  Passage 3: \{passage3\}
}
\end{tcolorbox}
\caption{Prompt for reader task.}
\label{fig:reader_prompt}
\end{figure}

Figure~\ref{fig:methodology_overview} shows an overview of our proposed pipeline. Starting with a question, we prompt the LLM to answer it (Figure~\ref{fig:answer_prompt}). We direct the LLM to act as an expert in order to set a more rigorous and less casual tone for the response~\cite{prompt_engineer}. Then, inspired by query expansion methods which have shown to be effective and help avoid topic drift problems \cite{carpineto2012survey,azad2019query,bodner1996knowledge,zighelnic2008query,amati2004query}, we combine the original question with the answer generated by the LLM for a second fact-checking or confirmation phase. We execute the combined query over a collection of passages to retrieve passages that are both relevant to the original question and that may support the LLM's generated answer. We then combine the original question, the generated answer, and the retrieved answer, and prompt the LLM to determine if the retrieved answer supports the generated answer (Figure~\ref{fig:stepped_classification}, Figure~\ref{fig:check_direct_answer_prompt}, and Figure~\ref{fig:validate_generated_answer_prompt}).
We summarize our proposed strategy as follows:

\begin{enumerate}
  \item Prompt the LLM to answer the question.
  \item Combine the LLM's answer with the original question.
  \item Execute the combined query on an external corpus (expected to contain correct answers), retrieving the most relevant passage(s).
  \item Prompt the LLM to compare its generated answer against the retrieved results from the combined query, with the goal of self-detecting hallucinations.
\end{enumerate}
In the following subsections, we elaborate on individual components of this pipeline.

\subsubsection{Retrieved Answer}
We experiment with three different types of retrieved answer:
\begin{itemize}
  \item BM25 retrieved answer: the most relevant passage retrieved using the Okapi BM25 ranking function.
  \item Neural retrieved answer: the most relevant passage retrieved using the multi-stage neural retrieval stack (SPLADE $+$ ANCE $+$ MonoT5 $+$ DuoT5).
  \item Reader extracted answer: obtained by prompting the LLM to act as a reader and extract a more concise answer from the top three passages retrieved by the multi-stage neural retrieval stack.
\end{itemize}

There can be multiple ways to address the same question, and the answers provided by the model may span various perspectives. Consequently, we experiment with retaining the top three retrieved passages. However, simply concatenating these passages without any refinement may lead to abrupt shifts, repetitions, or excessive length. Hence, we utilize a reader to perform a question-based summary of the top three retrieved passages, resulting in a summary that aligns better with the concise and direct nature of the LLM's generated answer. To do so, we prompt the LLM to act as a reader with the prompt shown in Figure~\ref{fig:reader_prompt}.

\subsubsection{Validating Generated Answer}
In order to validate generated answer against retrieved answer, we perform a stepped classification of each pair of question, generated answer, and retrieved answer as shown in Figure~\ref{fig:stepped_classification}. The first step is prompting the LLM with the prompt shown in Figure~\ref{fig:check_direct_answer_prompt} for it to decide if the generated answer and the retrieved answer actually address the question or not. In the second step, we classify the question, generated answer and retrieved answer pair. If the LLM classifies either one of the generated answer and the retrieved answer as not answering the question, we classify the question-answers pair as ‘Not Related’. Only if both answers are classified as direct answers to the question, we prompt the LLM to classify if the generated answer is supported by the retrieved answer with the prompt shown in Figure~\ref{fig:validate_generated_answer_prompt}.

We categorize the outcome of the LLM's decision into three different classes: 
\begin{itemize}
    \item We interpret the ``Yes'' class as indicating that there is no hallucination since the retrieved passages provide supporting evidence to the LLM's answer;
    \item We interpret the ``No'' class as indicating there is likely hallucination since the retrieved passages fail to support the LLM's answer;
    \item We interpret the ``Not Related'' class to indicate the LLM responded with a clarification request or claims it does not know the answer (For example, ``\emph{I would need more context to provide a specific answer...}'', ``\emph{I'm sorry, but I don't have access to ...}'' or ``\emph{I do not know. It's best to check ...}'') or the retriever failed to retrieve relevant passage. In either case, the question-answers pair will be irrelevant for the task of hallucination detection. 
\end{itemize}    

In addition, we also experiment with qrel passages, which are passages deemed highly relevant to the question by human annotators. By default, we assume these passages actually address the question without prompting the LLM with the prompt shown in Figure~\ref{fig:check_direct_answer_prompt}. Since these passages cannot be obtained free of human intervention, this is irrelevant to the main task of exploring the LLM's ability to automatically validate generated answers against a corpus. It is included to gain more insights into the effectiveness of our approach of combining the original question with the generated answer to curate a combined query for the fact-checking step.

\subsection{Results}

\subsubsection{Step 1: Classifying Answers}
\begin{table}
  \caption{LLM's classifications of the answers.}
  \label{tab:direct_answer_stats}
\centering
\scalebox{0.9}{
\begin{tabular}{p{0.04\linewidth} | >{\raggedright\arraybackslash}b{0.23\linewidth} >{\raggedright\arraybackslash}b{0.22\linewidth} >{\raggedright\arraybackslash}b{0.22\linewidth} >{\raggedright\arraybackslash}b{0.22\linewidth}}
\toprule
&\textbf{\small{Does the LLM’s generated \newline answer directly answer the \newline question?}}&\textbf{\small{Does the reader extracted \newline answer directly answer the \newline question?}}&\textbf{\small{Does the neural retrieved answer directly answer the question?}}&\textbf{\small{Does the BM25 retrieved answer directly answer the question?}}\\
\hline
\textbf{\small{Yes}} & \small{6,512 (93.30\%)} & \small{5,292 (75.82\%)} & \small{4,202 (60.20\%)} & \small{2,698 (38.65\%)} \\
\hline
\textbf{\small{No}} & \small{468 (6.70\%)} & \small{1,688 (24.18\%)} & \small{2,778 (39.80\%)} & \small{4,282 (61.35\%)} \\
\bottomrule
\end{tabular}
}
\end{table}

\begin{table}[]
\caption{Results of manually verifying LLM's classifications.}
\label{tab:direct_answer_manual_stats}
\centering
  \scalebox{0.95}{
\begin{tabular}{>{\raggedright\arraybackslash}b{0.17\linewidth} | b{0.13\linewidth} | >{\raggedright\arraybackslash}b{0.14\linewidth} >{\raggedright\arraybackslash}b{0.14\linewidth} >{\raggedright\arraybackslash}b{0.14\linewidth} >{\raggedright\arraybackslash}b{0.14\linewidth}}
\toprule
\textbf{\small{LLM's \newline Classification}} & \textbf{\small{Labeller's Opinion}} & \textbf{\small{Does the LLM’s generated answer directly answer the question?}} & \textbf{\small{Does the reader extracted answer directly answer the question?}} & \textbf{\small{Does the neural retrieved answer directly answer the question?}} & \textbf{\small{Does the BM25 retrieved answer directly answer the question?}} \\ 
\hline
\multirow{2}{*}{\parbox{2cm}{\textbf{\small{Yes}}}} & \textbf{\small{Correct}} & 98 & 99 & 99 & 92 \\
& \textbf{\small{Incorrect}} & 2 & 1 & 1 & 8 \\
\hline
\multirow{2}{*}{\parbox{2cm}{\textbf{\small{No}}}} & \textbf{\small{Correct}} & 82 & 58 & 38 & 62 \\
& \textbf{\small{Incorrect}} & 18 & 42 & 62 & 38 \\
\bottomrule
\end{tabular}}
\end{table}

Table~\ref{tab:direct_answer_stats} shows how LLM classifies different types of answers. According to the LLM's classifications, the generated answer provided by LLM adequately addresses the question in the majority of cases (approximately 93\% of the time). The order of relevance for the retrieved answers, from most to least relevant to the question, is as follows: the reader-extracted answer, the neural retrieved answer, and the BM25 retrieved answer, which aligns with the expectation.

To further investigate the reliability of these classifications, we manually examine and carefully inspect 100 randomly selected samples from each cell of Table~\ref{tab:direct_answer_stats}. In other words, for each of the four types of answers (generated answer, reader extracted answer, neural retrieved answer, and BM25 retrieved answer), we randomly select 100 samples where the LLM classified as direct answers and 100 samples where the LLM classified as non-direct answers. The manual inspection results are shown in Table~\ref{tab:direct_answer_manual_stats}.

From Table~\ref{tab:direct_answer_manual_stats} we see that in general there are far more false negatives than false positives. Furthermore, for the samples where the LLM classified as non-direct answers, its only achieves high accuracy for the generated answer but is only slightly better than random guesses (50\%) for all types of retrieved answers. This suggests that perhaps different prompt is needed for classifying the retrieved answers. 

In the following subsections, we will present specific instances and interesting patterns observed where the labeller and the LLM disagree.          
\paragraph{Classify Generated Answer}
\begin{figure}[!t]
\begin{tcolorbox}[colback=gray!5!white,colframe=gray!75!black]
\small{
    \textbf{Question}: what is the weather in powell wy

    \textbf{Generated Answer}: I'm sorry, as an AI language model, I don't have access to real-time weather information. However, you can check current weather conditions in Powell, WY by searching online or checking a weather app.

    \textbf{LLM’s Classification}: Yes, the answer directly answers the question by acknowledging its limitations and offering an alternative solution to finding the information requested.
    }
\end{tcolorbox}
\caption{False positive example of generated answer classification.}
\label{fig:generated_answer_false_positive}
\end{figure}

An illustrative example of false positive errors when the LLM classifies generated answer can be observed in Figure~\ref{fig:generated_answer_false_positive}, despite the prompt clearly stating that ``\emph{Examples of non-direct answers would be ... and provide some alternative ways to find answers}'', the LLM still makes an incorrect classification and explains its choice by ``\emph{the answer directly answers the question by acknowledging its limitations and offering an alternative solution}''. This example demonstrates that such errors cannot be rectified by simply stating the desired outcome in the prompt. Instead, better prompt engineering or fine-tuning of the LLM may be needed.

\paragraph{Classify Reader Extracted Answer}
\begin{figure}[!t]
\begin{tcolorbox}[colback=gray!5!white,colframe=gray!75!black]
\small{
    \textbf{Question}: who sang delta dawn?

    \textbf{Reader Extracted Answer}: Summary: Delta Dawn was originally recorded by Alexander Harvey in 1972 and later became a country hit for Tanya Tucker. Helen Reddy also had a number-one hit with the song in 1973.

    \textbf{LLM’s Classification}: No. The answer does not directly answer the question "who sang delta dawn?" Instead, it provides information about multiple artists who have recorded the song "Delta Dawn." While it does mention Tanya Tucker and Helen Reddy, who both had hits with the song, it doesn't specifically state who originally sang it.
    }
\end{tcolorbox}
\caption{Example of LLM's failures to identify synonymous expressions.}
\label{fig:extracted_answer_synonymous_failure}
\end{figure}

When classifying reader extracted answers, we observe a common type of mistakes where the LLM fails to recognize synonymous expressions as shown in Figure~\ref{fig:extracted_answer_synonymous_failure}. In the example, the LLM fails to recognize that ``someone sangs a song'' is the same as ``a song is recorded by someone''. Or, it might be overly stringent in demanding the exact terms to be provided in the response to consider it as a direct answer.

\paragraph{Classify Neural or BM25 Retrieved Answer}
\begin{figure}[!t]
\begin{tcolorbox}[colback=gray!5!white,colframe=gray!75!black]
\small{
    \textbf{Question}: which function automatically counts cells that meet multiple conditions

    \textbf{Retrieved Passage}: To count cells that aren't blank, use the COUNTA function. To count cells using multiple criteria, use the COUNTIFS function. See a video on how to use the COUNTIFS function. The SUMIF function adds only the values that meet a single criteria. The SUMIFS function adds only the values that meet multiple criteria

    \textbf{LLM’s Classification}: No, the answer does not directly answer the question. While the answer does mention a function that counts cells based on multiple conditions (COUNTIFS), it also provides information about other functions like COUNTA, SUMIF, and SUMIFS that are not relevant to the question. The answer should have simply stated that the function which automatically counts cells that meet multiple conditions is the COUNTIFS function.
    }
\end{tcolorbox}
\caption{Example of error caused by excessive details in retrieved passage.}
\label{fig:excessive_detail_retrieved_passages}
\end{figure}

\begin{figure}[!t]
\begin{tcolorbox}[colback=gray!5!white,colframe=gray!75!black]
\small{
    \textbf{Question}: hp envy size

    \textbf{Retrieved Passage}: New HP Envy laptops go big. Along with 14 and 15-inch models, the revamped Envy line highlights a rare new 17-inch laptop. [MUSIC] Wide variety of new high end back to school laptops, I'm Dan Ackerman, and this the new HP Envy line for 2015. Envy is the higher end of HP's PC line, and here we've got New 14 inch, 15 inch, and 17 inch models. They're all going to be available with current Gen Intel or AMD processors.
    \textbf{LLM’s Classification}: No. The answer does not directly answer the question about the size of the HP Envy. Instead, it provides information about the availability of different models and processors without specifying the exact size of the HP Envy laptop.
    }
\end{tcolorbox}
\caption{Example of the LLM fails to recognize implied information in retrieved passage.}
\label{fig:implied_info_retrieved_passages}
\end{figure}

When manually investigating classifications of retrieved passages, we observe that many false negative errors are caused by excessive details in the retrieved passages. As shown in Figure~\ref{fig:excessive_detail_retrieved_passages}, the LLM justifies its classifications by acknowledging that the passage does provide relevant information, but they also provide information about other functions or characteristics not asked by the question, thus does not directly answer the question. In other words, the LLM interprets ``direct'' as straight to the point and no other information, but we intend it to mean it does not provide alternative ways to find the answer. Even though our intention is explained in the prompt, more sophisticated prompt-engineering or choice of word is perhaps needed. It is natural for passages from the MS MARCO (V1) dataset to include additional detail than what the question asks for, since none of the passages is tailored to specific questions. Similarly, in reality, the evidence or supporting materials should frequently encompass more information than the claim being verified. This might be fixed by using a separate prompt to classify the retrieved passages, or using a reader to extract more concise answers from the passages before making classifications. 

Another common type of observed error is that the LLM often fails to recognize the answer is implied, as shown in Figure~\ref{fig:implied_info_retrieved_passages}. In particular, the LLM fails to recognize HP Envy laptops having ``14 inch, 15 inch, and 17 inch model'' means the possible sizes of HP Envy is 14, 15 and 17 inch. This implies that there is still potential for improvement in the LLM's language comprehension ability.
 
\subsubsection{Step 2: Classifying Question-answers Pair}
\begin{table}
  \caption{LLM's classifications of the question-answers pairs.}
  \label{tab:qa_pairs_classfications}
\centering
\scalebox{0.80}{
\begin{tabular}{b{0.17\linewidth} | >{\raggedright\arraybackslash}b{0.22\linewidth} >{\raggedright\arraybackslash}b{0.22\linewidth} >{\raggedright\arraybackslash}b{0.22\linewidth} >{\raggedright\arraybackslash}b{0.22\linewidth}}
\toprule
&\textbf{\small{Does the reader extracted answer support the LLM’s generated answer?}}&\textbf{\small{Does the neural answer support the LLM’s generated answer?}}&\textbf{\small{Does the BM25 answer support the LLM’s generated answer?}}&\textbf{\small{Does the qrel answer support the LLM’s generated answer?}}\\
\hline
\textbf{\small{Yes}} & \small{4,703 (92.02\%)} & \small{3,794 (93.93\%)} & \small{2,535 (95.62\%)} & \small{5,465 (83.92\%)} \\
\hline
\textbf{\small{No}} & \small{408 (7.98\%)} & \small{245 (6.07\%)} & \small{116 (4.38\%)} & \small{1,047 (16.08\%)} \\
\hline
\textbf{\small{Not Related}} & \small{1,869}	& \small{2,941} & \small{4,329} & \small{468} \\
\bottomrule
\end{tabular}}
\end{table}

\begin{table}[]
\caption{Results of manually verifying LLM's classifications of the question-answers pairs.}
\label{tab:qa_pairs_manual_classfications}
\centering
\scalebox{0.93}{
\begin{tabular}{>{\raggedright\arraybackslash}b{0.17\linewidth} | b{0.13\linewidth} | >{\raggedright\arraybackslash}b{0.15\linewidth} >{\raggedright\arraybackslash}b{0.15\linewidth} >{\raggedright\arraybackslash}b{0.15\linewidth} >{\raggedright\arraybackslash}b{0.15\linewidth}}
\toprule
\textbf{\small{LLM's \newline Classification}} & \textbf{\small{Labeller's Opinion}} & \textbf{\small{Does the reader extracted answer support the LLM’s generated answer?}} & \textbf{\small{Does the neural answer support the LLM’s generated answer?}} & \textbf{\small{Does the BM25 answer support the LLM’s generated answer?}} & \textbf{\small{Does the qrel answer support the LLM’s generated answer?}} \\ 
\hline
\multirow{2}{*}{\parbox{2cm}{\textbf{\small{Yes}}}} & \textbf{\small{Correct}} & 85 & 82 & 77 & 63 \\
& \textbf{\small{Incorrect}} & 15 & 18 & 23 & 37 \\
\hline
\multirow{2}{*}{\parbox{2cm}{\textbf{\small{No}}}} & \textbf{\small{Correct}} & 96 & 90 & 89 & 92 \\
& \textbf{\small{Incorrect}} & 4 & 10 & 11 & 8 \\
\bottomrule
\end{tabular}}
\end{table}

One of our main objectives is to investigate how many of the LLM’s answers suffer from hallucinations. Table~\ref{tab:qa_pairs_classfications} shows how LLM classifies its own answer against evidence. All of the percentages shown in Table~\ref{tab:qa_pairs_classfications} have excluded the ``Not Related'' cases as we interpret the ``Not Related'' class as not relevant for the task of validating the generated answer. Overall, after excluding the ``Not Related'', the LLM asserts that the retrieved material supports its own answer for about 93\% of questions. 

To further investigate the reliability of these classifications, we manually examine and carefully inspect 100 randomly selected samples from each cell in the first two rows of Table~\ref{tab:qa_pairs_classfications}. In other words, for each of the four types of answers that we compared the generated answer against (reader extracted answer, neural retrieved answer, BM25 retrieved answer, and qrel answer), we randomly select 100 samples where the LLM claims the generated answer is supported and 100 samples where the LLM claims the generated answer is not supported. The manual inspection results are shown in Table~\ref{tab:qa_pairs_manual_classfications}.

First of all, we observed that the qrel passages are not as reliable as sources of evidence for validating the generated answer, as they often present varying perspectives compared to the generated answer. This is also evident from Table~\ref{tab:qa_pairs_classfications} and Table~\ref{tab:qa_pairs_manual_classfications}, which shows that the qrel passages tend to have worse performance. This observation is in favor of our approach of using both the original question and the generated answer to curate a combined query for fact-checking.

\begin{figure}[!t]
\begin{tcolorbox}[colback=gray!5!white,colframe=gray!75!black]
\small{
    \textbf{Question}: how did van gogh paint

    \textbf{LLM's Answer}: Vincent van Gogh painted in a unique style characterized by bold brushstrokes, strong colors, and thick impasto. He often painted outdoors (en plein air) and used his emotions and personal experiences as inspiration for his work.

    \textbf{Evidence}: Vincent Van Gogh painted mainly using oil paint on canvas and was famous for his thick use of paint, adhering to the impasto technique which involves laying down a thick layer of paint in a particular segment of the canvas to make brushstrokes more visible and add texture.
    }
\end{tcolorbox}
\caption{Example of partially supported generated answer.}
\label{fig:validate_generated_partially_supported}
\end{figure}

We observed an average accuracy of about 80\% for samples that the LLM classified as ``Yes'' and 90\% for samples that the LLM classified as ``No''. In all cases, we observed slightly more false positives than false negatives. 

The primary reason behind these false positives is that the generated response includes many different assertions, and only a portion of them are supported by the retrieved answer. As shown in Figure~\ref{fig:validate_generated_partially_supported}, the ``\emph{thick impasto}'' claim is certainly supported by the retrieved answer, but ``\emph{strong colors}'', ``\emph{often painted outdoors}'', and ``\emph{used his emotions and personal experiences as inspiration}'' are not supported by the retrieved answer. Although the misclassification of partially supported answer may be solved by prompt engineering, it will not help finding evidence that covers more aspect of the generated answer. This observation leads to our second experiment discussed later in Section~\ref{sec:exp2}. 

\begin{figure}[!t]
\begin{tcolorbox}[colback=gray!5!white,colframe=gray!75!black]
\small{
    \textbf{Question}: population of bartholomew county indiana

    \textbf{LLM's Answer}: As per the US Census Bureau, the estimated population of Bartholomew County, Indiana as of 2019 is 83,779.

    \textbf{Evidence}: The population of Bartholomew County, Indiana as of 2015 is 81,162.

    \textbf{LLM's Classification}: Yes.  Explanation: The evidence provided states the estimated population of Bartholomew County, Indiana as of 2015 is 81,162. Since the answer states that the estimated population as of 2019 is 83,779, and the evidence supports that the population in 2015 was smaller, it indicates an increase in population over time. Therefore, the evidence supports the given answer.
    \newline
     \hrule 
    %\\
    \vspace{1em}
    \textbf{Question}: how old is bill gates wife

    \textbf{LLM's Answer}: Melinda Gates was born on August 15, 1964. As of 2021, she is 57 years old.

    \textbf{Evidence}: As of 2010, Melinda Gates, Bill Gates' wife and co-chair of the Bill and Melinda Gates foundation, is 46 years old. She was born in Dallas, Texas in the United States of America. Please see the related links for details.
    
    \textbf{LLM's Classification}: No. The evidence provided does not support the answer. The evidence states that as of 2010, Melinda Gates was 46 years old. However, it does not provide any information about her current age or indicate that she is 57 years old in 2021.
    }
\end{tcolorbox}
\caption{Examples of generated answer and supporting evidence from disparate time frames.}
\label{fig:question_answers_different_time}
\end{figure}

Another commonly observed type of error seems to arise from the time disparity between the answer and the evidence. As shown in Figure~\ref{fig:question_answers_different_time}, the population in 2015 says nothing about the population in 2019, unless a percentage increase or decrease is also provided. However, the LLM mistakenly interprets this as implicitly indicating an increase in population over time, when the evidence did not provide any support for it. On the other hand, one should be able to infer the age of someone in 2021 given their age in 2010. However, the LLM failed to recognize such connection. The LLM struggles to differentiate between what can be inferred and what cannot be inferred from the evidence from a different point in time than the answer.

\begin{figure}[!t]
\begin{tcolorbox}[colback=gray!5!white,colframe=gray!75!black]
\small{
    \textbf{Question}: anchor definition in literature

    \textbf{LLM's Answer}: In literature, an anchor is a recurring symbol or object that represents a character's values, beliefs, or memories. It is used to ground the character and provide continuity throughout the story.

    \textbf{Evidence}: Summary:   An anchor, usually made of metal, is a device attached to a ship or boat by a cable and cast overboard to hold it in a specific location. It has a fluke that digs into the bottom for stability. In literature, an anchor can also refer to a reliable or principal support, serving as a mainstay or backbone, providing emotional stability and support.

    \textbf{LLM's Classification}: Yes.  Explanation: The evidence provided explains the definition of an anchor in the context of literature. It mentions that an anchor in literature is a recurring symbol or object that represents a character's values, beliefs, or memories. It also states that it is used to ground the character and provide continuity throughout the story. This aligns with the given answer, which accurately describes what an anchor means in literature.
    }
\end{tcolorbox}
\caption{Example of the LLM confuses the generated answer with the evidence.}
\label{fig:confused_answer_evidence}
\end{figure}

Other common forms of error seem to be caused by the LLM confusing the answer with the evidence. As shown in Figure~\ref{fig:confused_answer_evidence}, LLM thinks that the evidence mentions ``\emph{an anchor in literature is a recurring symbol or object that represents a character's values, beliefs, or memories.}'' and ``\emph{is used to ground the character and provide continuity throughout the story.}'', while both of these are the exact phrases used in the answer. In the LLM's justification of its classification, it sometimes mistakenly took what is in the answer as also appearing in the evidence, despite it not being the case. This may be considered a form of LLM's hallucination, or that it is bad at clearly defining the boundary between the answer and the evidence in the input.

\section{Experiment 2} \label{sec:exp2}
Labellers observed that the generated answer usually contains multiple factual claims, and that a single piece of retrieved passage usually fails to address all of these aspects (Figure ~\ref{fig:validate_generated_partially_supported}). Therefore, we attempt to examine the generated answer at a more granular level in our second experiment. Inspired by FActScore~\cite{FactScore} and RARR~\cite{RARR}, we propose to break the generated answer into a list of factual statements, and then prompt the LLM to validate and post-edit each statement separately.

\subsection{Methodology}

\begin{figure}[]
  \centering
  \scalebox{0.8}{
  \includegraphics[width=\linewidth]{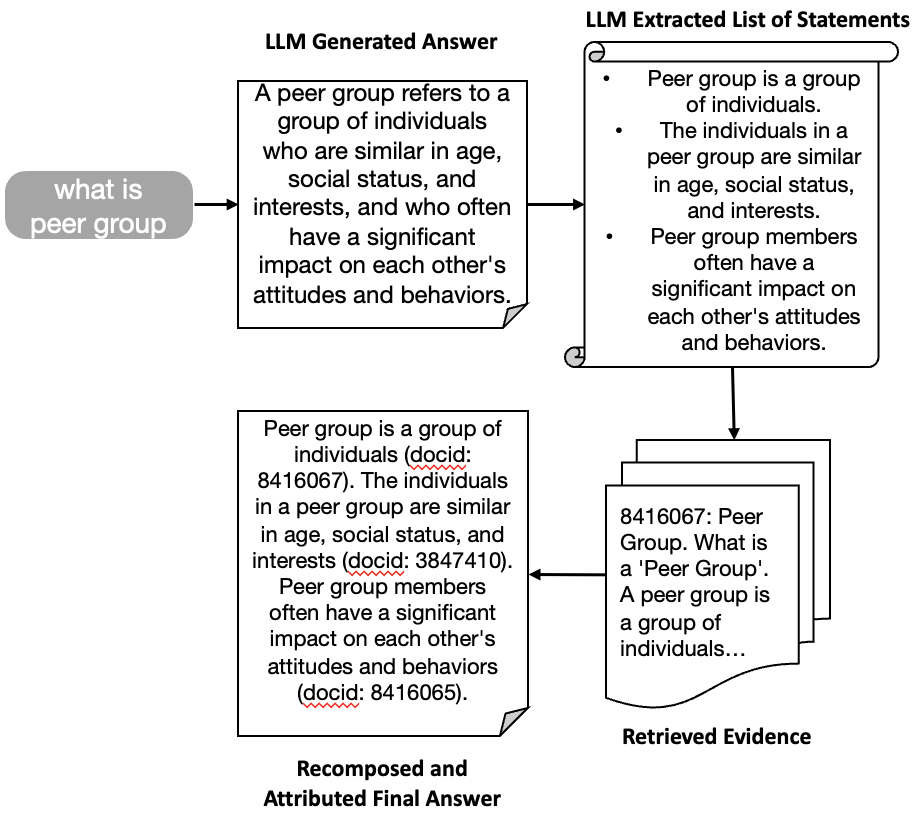}}
  \caption{Example of fact-based self-detecting hallucination in LLMs.}
  \label{fig:fact_methodology_example}
\end{figure}

 Figure~\ref{fig:fact_methodology_overview} shows an overview of our proposed pipeline, and Figure~\ref{fig:fact_methodology_example} shows an example. Starting with a question, we prompt the LLM to answer it (Figure~\ref{fig:answer_prompt}). Instead of directly verifying this generated answer as in the previous experiment, we ask the LLM to extract from the generated answer a list of factual statements worth validating in the context of the question (Figure~\ref{fig:fact_extract_prompt}). For each factual statement, we execute it over the collection of passages to retrieve passages relevant to the statement. We then prompt the LLM to validate and correct each factual statement using its corresponding retrieved evidence (Figure~\ref{fig:fact_validate_prompt} and Figure~\ref{fig:fact_post_edit_prompt}). In the end, we can recompose a final answer, which ideally would be free of hallucinations, with each assertion attributed to its supporting evidence. We summarize our strategy as follows:

\begin{figure}[]
  \centering
  \includegraphics[width=\linewidth]{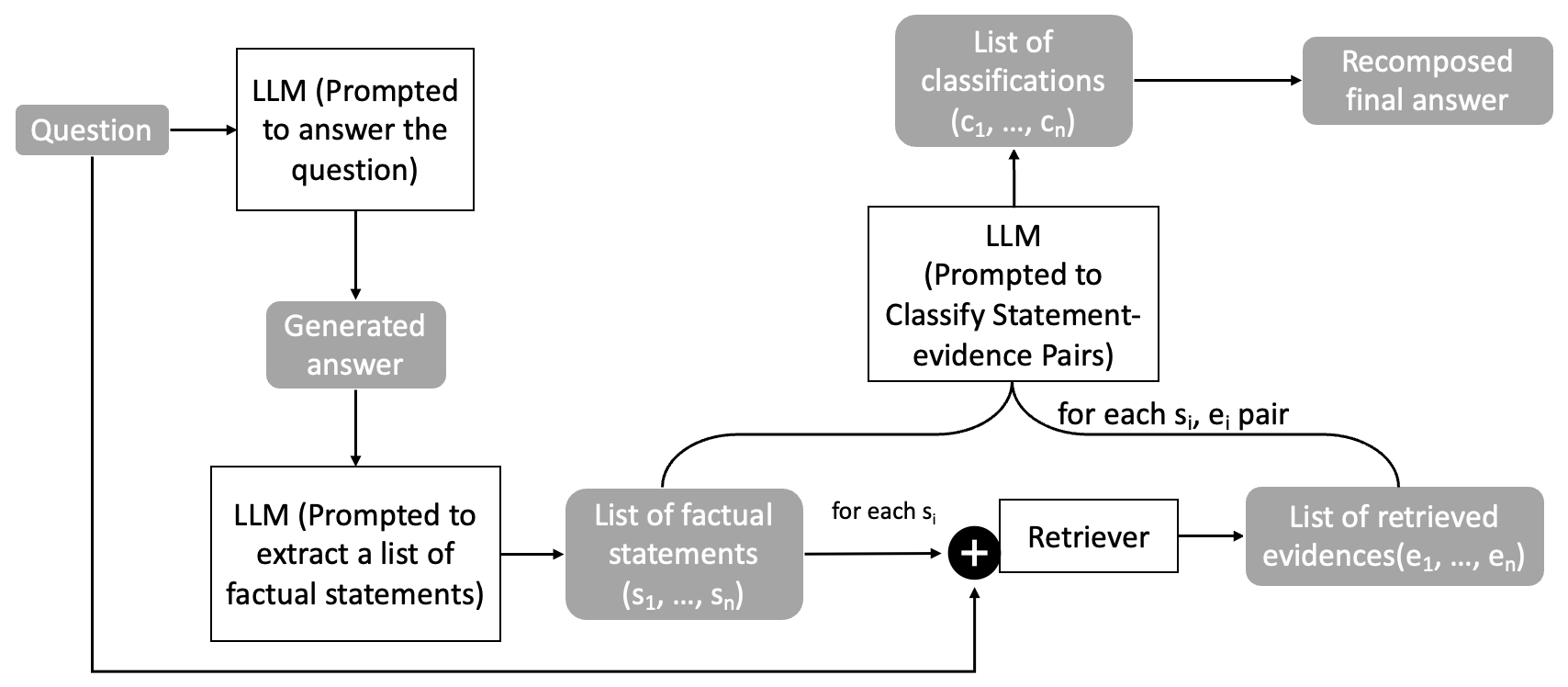}
  \caption{Overview of fact-based self-detecting hallucination in LLMs.}
  \label{fig:fact_methodology_overview}
\end{figure}

\begin{figure}[t]
\begin{tcolorbox}[colback=gray!5!white,colframe=gray!75!black]
\small{
 I want you to act as a language expert. Your task is given a question and a proposed answer, extract concise and relevant factual statements from the proposed answer. Include only statements that have a truth value and are worth validating, and ignore subjective claims. You should generate a bullet list of statements that are potentially true or false based on the question and proposed answer. Please only reply with the bullet list and nothing else.
  \newline
  \newline
 Question: \{question\}
 
 Proposed Answer: \{proposed answer\}
 }
\end{tcolorbox}
\caption{Prompt for extracting factual statements.}
\label{fig:fact_extract_prompt}
\end{figure}

\begin{figure}[t]
\begin{tcolorbox}[colback=gray!5!white,colframe=gray!75!black]
\small{
 I want you to act as a language expert and assist in determining the relationship between a factual statement and a piece of evidence. Here's how you should handle it:
 If the evidence supports the statement, reply with only the word 'Supported'.
 If the evidence contradicts the statement, reply with only the word 'Contradictory'.
 If the evidence is not relevant to the statement (neither supports nor contradicts it), reply with only the word 'Neither'.
 Your response should be a simple label 'Supported', 'Contradictory', or 'Neither', followed by a short and concise explanation of your classification.
  \newline
  \newline
 Statement: \{statement\}
 
 Evidence: \{passage\}
 }
\end{tcolorbox}
\caption{Prompt for validating a factual statement.}
\label{fig:fact_validate_prompt}
\end{figure}

\begin{figure}[t]
\begin{tcolorbox}[colback=gray!5!white,colframe=gray!75!black]
\small{
  I want you to act as a language expert and assist in post editing a false statement using a given piece of evidence. Your objective is to make minimal changes to the original statement while correcting it. Be concise. If the original false statement is one sentence, your corrected statement should also only be one sentence. Do not add more facts to the original statement, but only correct the wrong part of the original false statement. Please only reply with the corrected statement and nothing else.
  \newline
  \newline
  Statement: \{statement\}
  
  Evidence: \{passage\}
  }
\end{tcolorbox}
\caption{Prompt for post-editing factual statement.}
\label{fig:fact_post_edit_prompt}
\end{figure}

\begin{enumerate}
  \item Prompt the LLM to answer the question.
  \item Prompt the LLM to extract a list of factual statements from the LLM's answer. 
  \item Combine each factual statement with the original question. Execute the combined query on an external corpus, one at a time, retrieving the most relevant passage.
  \item Prompt the LLM to validate each factual statement against the retrieved passage, with the goal of self-detecting and self-correcting hallucinations.
\end{enumerate}

To validate and correct factual statements, we first prompt the LLM with the prompt shown in Figure~\ref{fig:fact_validate_prompt} and ask the LLM to decide if the factual statement is supported by the retrieved evidence. We categorize the outcome of the LLM's decision into three different classes: 
\begin{itemize}
    \item We interpret the ``Supported'' class as indicating there is no hallucination since the retrieved passage provides supporting evidence to the factual statement;
    \item We interpret the ``Contradictory'' class as indicating there is hallucination since the retrieved passage contradicts the factual statement;
    \item We interpret the ``Neither'' class as indicating the retrieved passage is not close enough to the factual statement for one to draw a definitive conclusion about whether the factual statement is supported or contradicted by the retrieved passage. 
\end{itemize} 
If a statement-evidence pair is classified as ``Contradictory'', we then prompt the LLM to post-edit the statement using the retrieved evidence with the prompt shown in Figure~\ref{fig:fact_post_edit_prompt}. Assuming everything works out, one can recompose a final answer free of hallucination and each factual claim in it can be linked to supporting evidence. This final answer can be constructed by simply concatenating all ``Supported'' statements and post-edited version of all ``Contradictory'' statements, and linking each statement to its corresponding retrieved passage.

\subsection{Results}
\begin{figure}[!t]
\begin{tcolorbox}[colback=gray!5!white,colframe=gray!75!black]
\small{
    \textbf{Question}: what are the monuments in washington dc

    \textbf{LLM Answer}: Washington DC is home to several monuments including the Washington Monument, Lincoln Memorial, Jefferson Memorial, ...

    \textbf{LLM's Extracted Factual Statements}: 
    \begin{itemize}
        \item Washington DC is home to the Washington Monument.
        \item Washington DC is home to the Lincoln Memorial.
        \item Washington DC is home to the Jefferson Memorial.
        ...
    \end{itemize}    
    }
\end{tcolorbox}
\caption{Example of helpful extraction of factual statements. Extraction of factual statements is particularly valuable when the LLM answer contains a list.}
\label{fig:good_fact_extract}
\end{figure}

\begin{figure}[!t]
\begin{tcolorbox}[colback=gray!5!white,colframe=gray!75!black]
\small{
    \textbf{Question}: who sings the song rise up

    \textbf{LLM's Answer}: Andra Day sings the song "Rise Up".

    \textbf{LLM's Extracted Factual Statements}: 
    \begin{itemize}
        \item The song "Rise Up" exists.
        \item There is a singer named Andra Day.
        \item Andra Day performs the song "Rise Up".
    \end{itemize}
    }
\end{tcolorbox}
\caption{Examples of extraction of factual statements at an excessively detailed and repetitive level.}
\label{fig:too_much_fact_extract}
\end{figure}

First of all, we examined the quality of the list of factual statements extracted. Out of the 6980 question-answer pairs, the LLM failed to extract any factual statement at all from 61 of them (similar to the ``Not Related'' class in the first experiment, where the LLM did not provide a direct answer).

The LLM generally decomposes the generated answer in an useful manner. For example, as shown in Figure~\ref{fig:good_fact_extract}, one piece of retrieved material likely would not contain exactly this list of monuments, so it can be helpful to validate each monument separately. However, the LLM also tends to generate a list of factual statements that may appear too detailed, as shown in Figure~\ref{fig:too_much_fact_extract}. In the context of the question ``\emph{who sings the song rise up}'', ``\emph{The song `Rise Up' exists.}'' and ``\emph{There is a singer named Andra Day.}'' are unnecessary as they are already implied. The LLM still has room of improvement to achieve the optimal level of granularity when extracting the list of factual statements.

\begin{table}
  \caption{LLM's classifications of the statement-evidence pairs.}
  \label{tab:fact_class_counts}
\centering
\scalebox{0.9}{
\begin{tabular}{>{\raggedright\arraybackslash}b{0.50\linewidth} |  >{\raggedright\arraybackslash}b{0.26\linewidth} >{\raggedright\arraybackslash}b{0.26\linewidth}}

\toprule
    \textbf{\small{LLM's Classification}} & \textbf{\small{Neural method retrieved evidence}} & \textbf{\small{BM25 method retrieved evidence}} \\
\hline
    \textbf{\small{Supported}} & 20,990 (83.14\%) & 20,158 (79.85\%) \\
    \textbf{\small{Contradictory}} & 3,128 (12.39\%) & 3,241 (12.84\%) \\
    \textbf{\small{Neither}} & 1,128 (4.47\%) & 1,847 (7.32\%) \\
\hline
    \textbf{\small{Average \% supported per query}} & 81.73\% & 78.59\% \\
    \textbf{\small{Average \% contradictory per query}} & 13.93\% & 14.15\% \\
    \textbf{\small{Average \% neither per query}} & 4.33\% & 7.26\% \\
\hline
    \textbf{\small{\# of fully supported responses}} & 4,241 (61.29\%) & 3,726 (53.85\%) \\
    \textbf{\small{\# of none supported responses}} & 364 (5.26\%) & 373 (5.39\%) \\
    \textbf{\small{\# of none contradictory responses}} & 4,761 (68.81\%) & 4,595 (66.41\%) \\
\bottomrule

\end{tabular}}
\end{table}

\begin{table}[]
\caption{Results of manually verifying LLM's classifications of the statement-evidence pairs.}
\label{tab:fact_class_counts_manual}
\centering
\begin{tabular}{>{\raggedright\arraybackslash}b{0.20\linewidth} | b{0.14\linewidth} | >{\raggedright\arraybackslash}b{0.24\linewidth} >{\raggedright\arraybackslash}b{0.24\linewidth} }
\toprule
\textbf{\small{LLM's Classification}} & \textbf{\small{Labeller's Opinion}} & \textbf{\small{Neural method retrieved evidence}} & \textbf{\small{BM25 method retrieved evidence}} \\ 
\hline
\multirow{2}{*}{\textbf{\small{Supported}}} & \textbf{\small{Correct}} & 83 & 72 \\
& \textbf{\small{Incorrect}} & 17 & 28 \\
\hline
\multirow{2}{*}{\textbf{\small{Contradictory}}} & \textbf{\small{Correct}} & 61 & 48 \\
& \textbf{\small{Incorrect}} & 39 & 52 \\
\hline
\multirow{2}{*}{\textbf{\small{Neither}}} & \textbf{\small{Correct}} & 90 & 87 \\
& \textbf{\small{Incorrect}} & 10 & 13 \\
\bottomrule
\end{tabular}
\end{table}

One of our main objectives is to investigate the degree to which LLM generated answers suffer from hallucination. Table~\ref{tab:fact_class_counts} shows how LLM classifies its factual claims against retrieved evidence. Overall, after excluding the ``Neither'' cases where the retrieved passage cannot be used to determine whether the factual claim hallucinates or not, the LLM believes the retrieved material supports about 85\% of its claims. The average percentage of claims supported per query is also about 80\%. Our automatic pipeline did not discover any hallucination for about 70\% of generated responses. Furthermore, according to the LLM's classifications, the evidence retrieved using the neural method does appear to be slightly better than the evidence retrieved using the BM25 method.

To further investigate whether the LLM is capable of correctly validating a factual claim against retrieved evidence, we manually inspected 100 randomly selected statement-evidence pairs from each cell of Table~\ref{tab:fact_class_counts}. In other words, for each retrieval method (BM25 and Neural), we randomly select 100 samples where the LLM believes the claim is supported by the retrieved evidence, 100 samples where the LLM believes the claim is contradicted by the retrieved evidence, and 100 samples where the LLM believes the claim is neither supported nor contradicted by the retrieved evidence. The results are shown in Table~\ref{tab:fact_class_counts_manual}. From Table~\ref{tab:fact_class_counts_manual}, we observe that the LLM's classification can achieve an average accuracy of about 80\% using neural method retrieved evidence. However, it tends to perform noticeably worse for the samples that it predicted to be in the ``Contradictory'' class.

\begin{figure}[!t]
\begin{tcolorbox}[colback=gray!5!white,colframe=gray!75!black]
\small{
    \textbf{LLM’s Extracted Factual Statement}: A dentist in the United States can expect to earn between \$120,000 to \$200,000 per year on average.

    \textbf{Evidence}: Conclusion about forensic pathologist salary. The forensic pathologists in the United States usually earn from between \$75,000 – \$200,000 per year however an average annual pay ranges from around \$80,000 – \$120,000 per year. 1  Share on Facebook.2  Share on Twitter.

    \textbf{LLM's Classification}: contradictory.  the evidence provided contradicts the factual statement. the evidence discusses the salary range of forensic pathologists in the united states, which is different from the statement about dentists' earnings. therefore, the evidence is not relevant to the statement and contradicts it.
    }
\end{tcolorbox}
\caption{Example of falsely classifying ``Neither'' sample as ``Contradictory''.}
\label{fig:misclass_neither_to_cont}
\end{figure}

One common type of error observed is that the LLM falsely classified ``Neither'' sample as ``Contradictory'', and it explains its classification using the definition of the ``Neither'' class in the prompt. As shown in the example in Figure~\ref{fig:misclass_neither_to_cont}, the LLM recognizes that forensic pathologists' salary is irrelevant to dentists' salary, but it classifies this sample as ``Contradictory'' on the basis that ``the evidence is not relevant to the statement and contradicts it''. However, in the prompt (Figure~\ref{fig:fact_validate_prompt}), we clearly defined ``Neither'' as ``\emph{If
the evidence is not relevant to the statement (neither supports nor contradicts it),
reply with only the word `Neither'}''. 

Furthermore, it is noteworthy that although the LLM falsely classified many samples as ``Contradictory'', later in the post-editing phase it often returns the original statement. When the classification is correct, the post-edited statement is usually of reasonable quality.

\section{Limitations}

We recognize several important limitations of this research, specifically:
\begin{enumerate}[wide, labelwidth=!, labelindent=0pt]
\item
Our experiments used only a single language model, which we choose for its convenient and inexpensive API (gpt-3.5-turbo).
\item
We kept our prompts simple and natural, with minimal prompt engineering. We feel that excessive prompt engineering can harm the reproducibilty of the experiments. With simple and natural prompts, future language models could be expected to preform reasonably. Nonetheless, we follow what we perceive to be current ``best practice'' for example, by framing the context of requests by indicating that the LLM should act as an expert.
\item
The entirety of the MS MARCO collection, including all questions and passages, may have been included in the training data for the model. Given the size and scope of the training data for the OpenAI GPT models, we assume it has, but we do not know for sure.
\item
We chose MS MARCO because it is a relatively large collection of questions with a corpus known to contain answers. In future work, we plan to explore other benchmark collections. 
\item
All questions have answers in the corpus, although not necessarily the answers that are consistent with those generated by the LLM.
\end{enumerate}
Different models, including later generations of the GPT family, and additional prompt engineering may improve the ability to predict hallucinations. Theoretically, if the questions and answers are included in the training data, the LLM could recognize the questions and respond with answers based on the MS MARCO passages, reducing the potential for hallucinations. If the corpus and questions are included the training data for the LLM, and all questions are answered by the corpus, the current experiment may be viewed as a ``best case'' scenario.

\section{Conclusion}
In this paper, we investigate the LLM's ability to self-detect hallucinations in its generated texts  with the help of an information retrieval system to retrieve supporting evidence. The methodology we proposed in the first experiment (Figure~\ref{fig:methodology_overview}) is perhaps the simplest possible for this purpose. Based on observations made when manually labelling the data, we proposed another experiment (Figure~\ref{fig:fact_methodology_overview}). The second experiment aims to resolve the frequently occurring problem in the first experiment, which is the evidence only partially supports the texts to be validated. In addition, the second experiment further attempts to produce a final fully-attributed output free of hallucination. Generally, in over 80\% of cases, the LLM is able to verify its generated texts when provided with relevant supporting material. However, when we manually examine its decisions, we observed that the LLM sometimes behave unreasonably. For example, it acts contrary to the given prompt, fabricates evidence to support statement or answer, and misses obvious or implied connections. These observations opens up a room for further research in this area. Nevertheless, one cannot solely rely on this approach to detect hallucinations because the LLM is observed to make more false positive errors than false negative errors when checking if the generated answer is free of hallucination.

In the future, we plan to experiment with more prompts or train specific language model for each specific task. We may also experiment with different LLMs, especially those with access to predicted token probabilities. Overall, we believe that validation by retrieving supporting evidence has the potential to provide a simple and reliable solution for detecting and ameliorating LLM hallucinations.

\balance
\bibliographystyle{ACM-Reference-Format}
\bibliography{sample-base}
\balance


\end{document}